\documentclass[preprint]{imsart} 
\newcommand{\tr}{\mbox{tr}}
\newcommand\indep{\protect\mathpalette{\protect\independenT}{\perp}}

\usepackage[figuresright]{rotating}
\usepackage{amsmath}
\usepackage{amssymb}
\usepackage{subfigure}
\usepackage{pstricks,pst-node,pst-plot}
\usepackage{color}
\usepackage{pst-all}
\usepackage{pspicture}
\usepackage{booktabs}
\usepackage{wasysym}
\usepackage[all]{xy}
\xyoption{arc}
\usepackage{epsfig,graphicx}
\usepackage{theorem}
\usepackage{tikz}
\usetikzlibrary{positioning,arrows.meta,quotes}
\usetikzlibrary{shapes,backgrounds}
\usepackage{natbib}

\newcommand\indepen{\protect\mathpalette{\protect\independenT}{\perp}}
\def\independenT#1#2{\mathrel{\rlap{$#1#2$}\mkern5mu{#1#2}}}

\newcounter{count}

\begin{document}
\setcounter{count}{0}
\begin{frontmatter}

\title{Relations between networks, regression, partial correlation, and latent variable model}
\runtitle{Intervention in undirected graphs}

\author{Lourens Waldorp}
\author{Maarten Marsman}
\runauthor{Waldorp and Marsman}

\address{\tt\small waldorp@uva.nl\hspace{3em} m.marsman@uva.nl}

\begin{abstract}
The Gaussian graphical model (GGM) has become a popular tool for analyzing networks of psychological variables. In a recent paper in this journal, Forbes, Wright, Markon, and Krueger (FWMK) voiced the concern that GGMs that are estimated from partial correlations wrongfully remove the variance that is shared by its constituents. If true, this concern has grave consequences for the application of GGMs. Indeed, if partial correlations only capture the unique covariances, then the data that come from a unidimensional latent variable model ULVM should be associated with an empty network (no edges), as there are no unique covariances in a ULVM. We know that this cannot be true, which suggests that FWMK are missing something with their claim. We introduce a connection between the ULVM and the GGM and use that connection to prove that we find a fully-connected and not an empty network associated with a ULVM. We then use the relation between GGMs and linear regression to show that the partial correlation indeed does not remove the common variance. 
\end{abstract}

\end{frontmatter}


\section{Introduction}\noindent

\nocite{ForbesEtAl_IP_MBR}

In a recent paper in this journal, Forbes, Wright, Markon and Krueger \citeyearpar[][henceforth FWMK]{ForbesEtAl_IP_MBR} voiced the concern that Gaussian graphical models (GGMs) that are estimated from partial correlations wrongfully remove crucial information from the data: The variance that is shared by its constituents. This concern is fundamental to their evaluation of the use of network models in psychopathology \citep[see, for instance;][]{ForbesEtAl_2017, ForbesEtAl_2019_WorldPsychiatry}. FWMK are under the impression that if an edge between two variables is estimated using a partial correlation, ``the edge is based on the variance shared by each pair of [variables] \emph{after removing the variance they share with all other [variables] in the network} (p. 13, their italics).\footnote{Even though their concern is only about the use of network models in the context of psychopathology, their critique is about a statistical concept or method and not about context. We have therefore replaced the word "symptom" with "variable" in this quote to express the generality of their concern.}
When the network comprises many variables, a large part of the covariance between the two focal variables is shared with other variables in the network. As a result, the region of unique covariance between the focal variables shrinks with the size of the network, and estimated edges become ``unreliable'' and primarily made up of ``random and systematic error'' (p. 14). 
Here we show that the concerns of FWMK are wrong. 

We illustrate the concern of FWMK for a three-variable network in Figure \ref{fig:explained-variance}, which we will also use later in Section \ref{sec:partial-correlation-regression}. We aim to obtain the relation between the variables $X_1$ and $X_3$ at the population level. Their covariance comprises two parts; one part that embodies the variance that is shared only by the two variables $X_1$ and $X_3$, and one part that embodies the variance that is shared by the two variables $X_1$ and $X_3$ in conjunction with the other variable $X_2$. In Figure \ref{fig:explained-variance}(a), we see a representation of the overlap in variance between all three variables. 
In Figure \ref{fig:explained-variance}(b), we see the complete covariance between $X_1$ and $X_3$. In Figure \ref{fig:explained-variance}(c), on the other hand, the overlap between $X_1$ and $X_2$ has been removed (partialled out) and only considers the unique contributions of $X_1$ and $X_2$ on $X_3$. The concern of FWMK is that partial correlations remove the overlap between $X_1$ and $X_2$ in estimating the relation between $X_1$ and $X_3$. We will demonstrate, however, that  this view is incorrect. That is, all the variance of $X_3$ that $X_1$ and $X_2$ \textit{could} explain \textit{is} explained using partial correlations, and so no information is lost. 
We provide two arguments that show that no information is lost when using partial covariances (or regression coefficients) with respect to covariances. The first is that the partial covariance matrix is a one-to-one correspondence (i.e., a bijection) with the covariance matrix. This implies that you can go back and forth between the two worlds, and that these partial covariance and covariance worlds are basically the same. The second argument uses the regression perspective and shows that the explained variance ($R^{2}$) contains all shared variance from the predictors, and so nothing is lost.


\def\firstcircle{(0,0) circle (1cm)}
\def\secondcircle{(45:1.5cm) circle (1cm)}
\def\thirdcircle{(0:1.5cm) circle (1cm)}
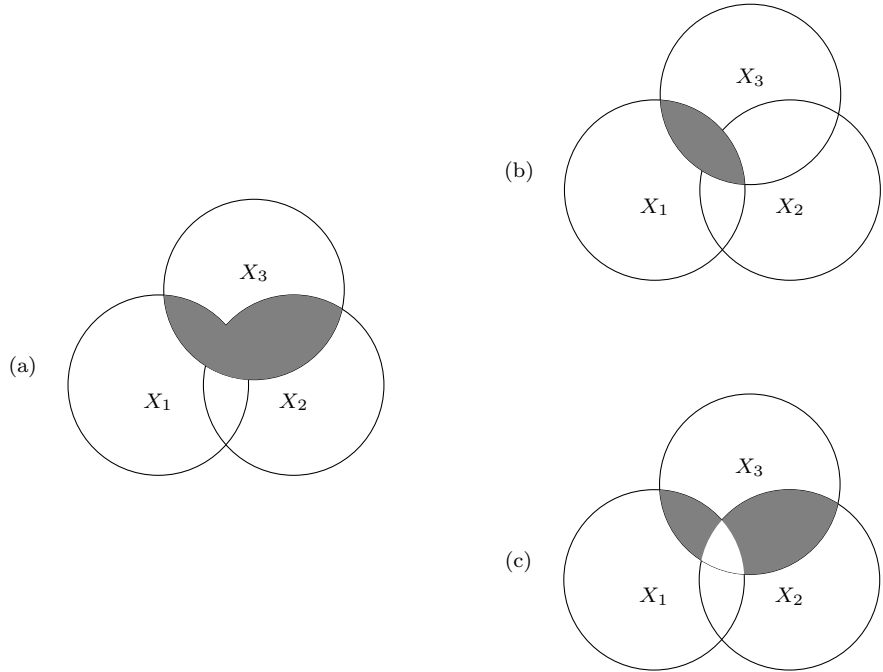
\begin{figure}[t]\centering
\begin{tabular}{ c @{\hspace{5em}} c }
&
\begin{tikzpicture}[scale = 1.2]
    \draw \firstcircle node[below] {$X_{1}$};
    \draw \secondcircle node [above] {$X_{3}$};
    \draw \thirdcircle node [below] {$X_{2}$};
    \draw  (-1.5,0) node [left,above] {(b)};
    \begin{scope}
      \clip \firstcircle;
      \fill[gray] \secondcircle;
    \end{scope}
\end{tikzpicture}
\\[-4em]
\begin{tikzpicture}[scale = 1.2]
    \draw \firstcircle node[below] {$X_{1}$};
    \draw \secondcircle node [above] {$X_{3}$};
    \draw \thirdcircle node [below] {$X_{2}$};
    \draw  (-1.5,0) node [left, above] {(a)};
    \begin{scope}
      \clip \firstcircle;
      \fill[gray] \secondcircle;
    \end{scope}
    \begin{scope}
      \clip \secondcircle;
      \fill[gray] \thirdcircle;
    \end{scope}
    \begin{scope}
      \clip \firstcircle;
      \clip \secondcircle;
      \fill[gray] \thirdcircle;
    \end{scope}
\end{tikzpicture}
&\\[-4em]
&
\begin{tikzpicture}[scale = 1.2]
    \draw \firstcircle node[below] {$X_{1}$};
    \draw \secondcircle node [above] {$X_{3}$};
    \draw \thirdcircle node [below] {$X_{2}$};
    \draw  (-1.5,0) node [left, above] {(c)};
    \begin{scope}
      \clip \firstcircle;
      \fill[gray] \secondcircle;
    \end{scope}
    \begin{scope}
      \clip \secondcircle;
      \fill[gray] \thirdcircle;
    \end{scope}
    \begin{scope}
      \clip \firstcircle;
      \clip \secondcircle;
      \fill[white] \thirdcircle;
    \end{scope}
\end{tikzpicture}
\end{tabular}
\caption{
A network of three variables (variables are nodes) or the regression of node $X_{3}$ on the predictors $X_{1}$ and $X_{2}$. Part (a) shows all variance that the predictors share with the dependent variable. Part (b) shows the contribution of $X_1$ to the explained variance in regression (i.e., $R^{2}$). Part (c) illustrates the variance that comes from each regressor separately. The shared variance is removed from the contribution of the regressors to prevent bias in the associated coefficients.}
\label{fig:explained-variance}
\end{figure}

FWMK's conviction is that partial correlations only capture the unique covariances (the shared variance between $X_1$ and $X_3$ that does not overlap with $X_2$) but not the shared covariances (the shared variance between $X_1$, $X_2$, and $X_3$). 
If the partial correlation indeed excludes the shared covariances, then the data that come from a unidimensional latent variable model (ULVM) should be associated with an empty network (no edges), as there are no unique covariances in a ULVM. This result is crucial since we often use instruments that are consistent with low-dimensional or ULVM, such as IQ-tests, in psychology. If partial correlations indeed only use the unique parts of the covariances, networks based on the ULVM, data of IQ tests, for example, would at the population level be empty and contain no edges. As a result, the GGM is useless in the case of the ULVM, even at a descriptive level, as it is unable to convey the most basic observation in intelligence research, the positive manifold. We agree that if this view is correct, the future of GGMs applied to psychological data would be dire. 


It is hard to overstate the severity of the above conclusion about GGMs. However, it also suggests that its premises must be wrong, as it is well-known that if the data come from a unidimensional latent variable model, the estimated network is going to be fully-connected and not empty. That these networks are fully-connected was theoretically and empirically shown in the case of binary variables, using a connection between binary latent variable models and binary network models \citep{EpskampEtAl_2018_HoP, MarsmanEtAL_2018_MBR, MarsmanEtAl2015SciRep}, and was also proven in the general case by, for example, \citet{HollandRosenbaum1986}. At a minimum, this implies that there is an essential element missing in the understanding of GGMs and partial correlations, and this paper aims to fill that gap.


The remainder of this paper comprises three parts. In the first part, we formally introduce the ULVM and GGM and consider the role that partial correlations play in the estimation of a GGM. In the second part, we will analyze the theoretical relationship between the GGM and the ULVM and show that one indeed does expect to obtain a fully-connected network from data that come from the ULVM. In the third part, we revisit the relationship between linear regression, the GGM, and partial correlation to prove that the GGM estimated from partial correlations indeed conveys the shared variance. 

\section{Models}\noindent
In this section, we introduce the unidimensional latent variable model (ULVM) and the GGM. We show how the assumptions about the ULVM's regression from the latent variable to the observed variables leads to a particularly simple form of the population covariance matrix. We will use the expression that we obtain for the covariance matrix to relate the ULVM to the GGM in the next section. There, we will also show that for the ULVM, observed partial correlations would all be positive. That this proves our first point that the GGM applied to data coming from a ULVM  will be fully-connected and not empty, relies on the fact that estimating the edges in a GGM is equivalent to computing the nonzero elements in the matrix of partial correlations. We will show that determining the matrix of partial correlations is equivalent to obtaining the independence relations between variables in the network in this section, and provide a small example to illustrate the principle. 

\subsection{The Unidimensional Latent Variable Model}
\label{sec:latent-variable-model}
\begin{figure}[t]
\hspace{2em}
\begin{tikzpicture}[scale=1.1,
  transform shape, node distance=2cm,
  roundnode/.style={circle, draw=black, thick, minimum size=7mm},
  squarednode/.style={rectangle, draw=black, thick, minimum size=5mm},
  arrow/.style = {semithick,-Stealth},
  dotnode/.style={fill,inner sep=0pt,minimum size=2pt,circle} 
]

\node[roundnode] (latent) {$\eta$};
\node[squarednode, left=1.2cm of latent,yshift=0.5cm] (X2) {$X_2$}; 
\node[squarednode, above=0.5cm of X2]    (X1) {$X_1$};
\node[squarednode, below=0.5cm of X2]    (X3) {$X_3$};
\node[squarednode, below=0.5cm of X3]    (X4) {$X_4$};

\foreach \i/\Yshift in {1/0,2/-3pt,3/-2pt,4/0}
{
   \node[left=1cm of X\i] (e\i) {$e_\i$};
   \draw[arrow] (latent) -- node["$\lambda_\i$"{inner sep=1pt,yshift=\Yshift}]{} (X\i.east);
   \draw[arrow] (e\i) -- (X\i);
}


\end{tikzpicture}
\hspace{5em}~
\raisebox{2em}{
\begin{tikzpicture}[scale=1.1,
  transform shape, node distance=2cm,
  roundnode/.style={circle, draw=black, thick, minimum size=7mm},
  squarednode/.style={rectangle, draw=black, thick, minimum size=5mm},
  arrow/.style = {semithick,-Stealth},
  dotnode/.style={fill,inner sep=0pt,minimum size=2pt,circle} 
]

\node[roundnode] (X1) {$X_{1}$};
\node[roundnode, right=1.2cm of X1] (X2) {$X_{2}$};
\node[roundnode, below=1.2cm of X1] (X3) {$X_{3}$};
\node[roundnode, below=1.2cm of X2] (X4) {$X_{4}$};

 \draw[-, thick] (X1) -- node[above]{$\alpha\lambda_{1}\lambda_{2}$} (X2);
 \draw[-, thick] (X1) -- node[left]{$\alpha\lambda_{1}\lambda_{3}$} (X3);
 \draw[-, thick] (X1) -- node[left]{\tiny $\alpha\lambda_{1}\lambda_{4}$} (X4);
 \draw[-, thick] (X2) -- node[right]{\tiny $\alpha\lambda_{2}\lambda_{3}$} (X3);
 \draw[-, thick] (X2) -- node[right]{$\alpha\lambda_{2}\lambda_{4}$} (X4);
 \draw[-, thick] (X3) -- node[below]{$\alpha\lambda_{3}\lambda_{4}$} (X4);
\end{tikzpicture}}

\caption{The left panel shows a unidimensional latent variable model with observed variables $X_{1}, X_2, X_3$ and $X_{4}$, a scalar latent variable $\eta$, and loadings $\lambda_{1}, \lambda_2, \lambda_3$ and $\lambda_{4}$. The $e_{i}$ on the left are the error terms for the observed variables. 
The right panel shows the associated network model. All observed variables are connected to each other with parameter $\alpha\lambda_{i}\lambda_{j}$, where $-\alpha^{-1}=\lambda^{\sf T}\lambda+1$  (see (\ref{eq:inverse-covariance}) and further for details on the weights)}. 
\label{fig:latent-network}
\end{figure}
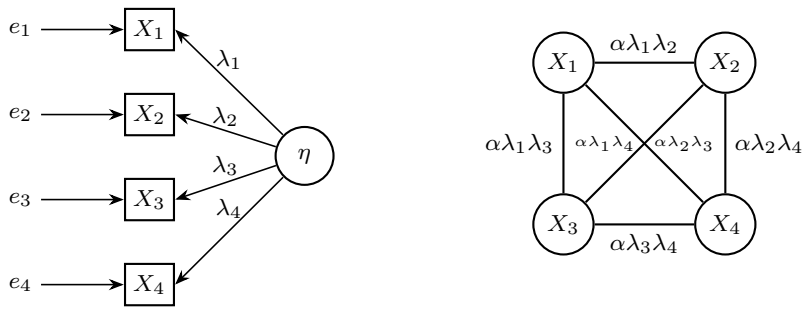

The ULVM assumes that there is a single latent variable $\eta$ (a random variable) that can explain away the correlations between the observed random variables $X_{1},X_{2},\ldots,X_{p}$. In other words, we have that $X_{i}$ and $X_{j}$ are independent conditional on the latent variable $\eta$. This conditional independence implies that the correlation between $X_{i}$ and $X_{j}$ is 0 given $\eta$. This assumption is often called local independence and is written as $X_{i}\indepen X_{j}\mid \eta$, where the symbol $\indepen$ stands for statistical independence \citep{Dawid1979}. The relation between each observed variable $X_{i}$ and the latent variable $\eta$ is often assumed linear, that is, 
\begin{align}
X_{i} = \lambda_{i}\eta + e_{i}
\end{align}
where $\lambda_{i}$ is the loading (regression coefficient) for the relation between the observed- and latent variables, and $e_{i}$ is the error (or residual if there is misspecification). See Figure \ref{fig:latent-network}, left panel, for a graphical illustration of the model. We assume that both the observed- and latent variables are continuous and have a joint Gaussian distribution. 

By considering covariances between the variables given the linear model above we find intuitive notions about what to expect from such a model. Suppose that the mean and variance of the latent variable $\eta$ are $\mu_{\eta}$ and $1$, respectively, and that the mean and variance of the error $e_{i}$ are $0$ and $1$, respectively. 
We assume that the errors of different variables are uncorrelated and that the errors are also uncorrelated with the latent variable. These assumptions are 
\begin{itemize}
\item[(a)] $\mathbb{E}(e_{i})=0$ and $\mathbb{E}(e_{i}e_{j})=0$, $\mathbb{E}(e_{i}^{2})=1$
\item[(b)] $\mathbb{E}[e_{i}(\eta-\mu_{\eta})]=0$ \hfill \refstepcounter{equation}\textup{(\theequation)}
\item[(c)] $\mathbb{E}(\eta)=\mu_{\eta}$ and $\mathbb{E}(\eta-\mu_{\eta})^{2}=1$
\end{itemize}
With these assumptions we find the following expression for the marginal covariance of variables $X_i$ and $X_j$, $i\ne j$,
\begin{align}
\text{cov}(X_{i},X_{j})&=\mathbb{E}(X_{i}-\lambda_{i}\mu_{\eta})(X_{j}-\lambda_{j}\mu_{\eta}) \notag\\
	&=\mathbb{E}(\lambda_{i}(\eta-\mu_{\eta})+e_{i})(\lambda_{j}(\eta-\mu_{\eta}) +e_{j})=\lambda_{i}\lambda_{j}.
\end{align}
If $i=j$, then we obtain $\text{var}(X_{i})=\lambda^{2}_{i}+1$. In other words, the covariance matrix of the random variables in the $p$ vector $\mathbf{x}=(X_{1},X_{2},\ldots,X_{p})^{\sf T}$ is equal to the $p\times p$ matrix 
\begin{align}
\boldsymbol{\Sigma}=\boldsymbol{\lambda}\boldsymbol{\lambda}^{\sf T}\mathbb{E}(\eta-\mu_{\eta})^{2}+\mathbb{E}(\mathbf{e}\mathbf{e}^{\sf T})=\boldsymbol{\lambda}\boldsymbol{\lambda}^{\sf T}+\mathbf{I}_{p}
\end{align}
where $\mathbf{I}_{p}$ is the $p\times p$ identity matrix and $\boldsymbol{\lambda}=(\lambda_{1},\ldots,\lambda_{p})^{\sf T}$ and $\mathbf{e}=(e_{1},\ldots, e_{p})^{\sf T}$ are $p$ vectors. In an empirical analysis the interest is in estimating the parameters $\boldsymbol{\lambda}$ by fitting this expected variance matrix to the sample variance matrix. 

When we condition on the latent variable $\eta$ we obviously obtain a different covariance matrix. 
We fix $\eta$ to any particular value (conditioning) and then determine expectations. 
We find the following expression for the conditional covariance of variables $X_i$ and $X_j$, $i\ne j$,
\begin{align}
\text{cov}(X_{i},X_{j}\mid \eta)=\mathbb{E}[(X_{i}-\lambda_{i}\eta)(X_{j}-\lambda_{j}\eta)\mid \eta]=\mathbb{E}(e_{i}e_{j}\mid \eta)=0
\end{align}
and the value $1$ if $i=j$. 
This shows that conditional on the latent variable $\eta$ the correlations between any of the observed variables is indeed equal to $0$. 
\subsection{The Gaussian Graphical Model}
\label{sec:gaussian-graphical-model}
What do we mean by a network or graphical model? In the case where all variables have a joint Gaussian (multivariate normal) density, we speak of a GGM. A GGM refers to the correspondence between a picture of a network and conditional independence relations. In particular, the nodes of the network $G=(V,E)$ in $V=\{1,2,\ldots,p\}$ are associated with random variables $X_{1},X_{2},\ldots,X_{p}$, and the edges of the network in $E=\{(i,j)\in V\times V: i - j\}$ indicate that whenever variables $i$ and $j$ are neighbours (adjacent), i.e., $i - j$, then $X_{i}$ is dependent on $X_{j}$ given all remaining variables $X_{V\backslash \{i,j\}}$, where the set $V\backslash \{i,j\}$ is the set of nodes $1,2,\ldots,p$ with the nodes $i$ and $j$ removed. For Gaussian random variables, it turns out that determining that two variables are independent given all other variables, is the same as checking if the partial correlation between these two variables is equal to $0$ \citep[][Section 5.1.3]{Lauritzen96}.  
It turns out that the matrix of partial covariances of all variables corresponds exactly to the inverse of the (co)variance matrix $\boldsymbol{\Sigma}$ of all variables $X_{1},X_{2},\ldots,X_{p}$. The partial correlations can be obtained from the matrix of partial covariances by dividing each off-diagonal element with the product of corresponding diagonal elements. The inverse $\boldsymbol{\Sigma}^{-1}=\boldsymbol{\Theta}$ is often referred to as the concentration matrix. So, in a  multivariate Gaussian distribution all we need do is to determine the zeros in the concentration matrix and we have found our conditional independencies. 
\citet[][Proposition 5.2]{Lauritzen96} showed that a zero in the concentration matrix corresponds to a conditional independence relation, i.e.,
\begin{align}
\Theta_{ij}=0	\quad\Longleftrightarrow\quad	X_{i}\indepen X_{j}\mid X_{V\backslash\{i,j\}}.
\end{align}
Note that we condition on all remaining variables in $V\backslash\{i,j\}$. And so an edge $i-j$ will be in the network $G$ if and only if $X_{i}$ is dependent on $X_{j}$ conditional on all other variables in $V\backslash\{i,j\}$. We could also say that, given the set of variables in $V$, we can find no alternative explanation for the dependence between $X_{i}$ and $X_{j}$ \citep{Pearl:2001}.

To illustrate the role of partial covariance (and correlation) in the GGM, we consider a small example with three nodes $V=\{1,2,3\}$ and two edges $E=\{1-2,1-3\}$. Suppose that we have the following variance matrix $\Sigma$ and concentration matrix $\Theta=\Sigma^{-1}$
\begin{align*}
\boldsymbol{\Sigma}=
\begin{pmatrix}
2	&-1		&-1\\
-1	&1.5		&0.5\\
-1	&0.5		&1.5
\end{pmatrix}
\quad
\text{and}
\quad
\boldsymbol{\Theta}=
\begin{pmatrix}
1	&0.5		&0.5\\
0.5	&1		&0\\
0.5	&0		&1
\end{pmatrix}
\end{align*}
We notice that the variables $X_{2}$ and $X_{3}$ have covariance 0.5 (correlation $\rho_{23}=0.5/1.5=\tfrac{1}{3}$) but are not correlated conditional on variable $X_{1}$ (partial correlation $\rho_{23\mid 1}=0$). The conditional independence can be interpreted as having found an alternative explanation for the correlation between variables $X_{2}$ and $X_{3}$, namely their relation to variable $X_{1}$. 

Thus, a GGM provides information on possible alternative explanations for correlations. In other words, if we find a zero partial correlation, then we know that there is no unique connection between the variables; if there is non-zero partial correlation, then we know that no other variables can explain away the obtained correlation.

\section{The Relation Between the GGM and ULVM}
\label{sec:relations-networks-latent}
An obvious question for any researcher considering both networks and latent variable models is: What are the similarities and how can I characterise them? Here we consider the case of a ULVM and determine what network corresponds to such a model. 
 That is, if a ULVM holds for the observed variables, then what does this imply for a network of only observed variables? 
The answer is that we would obtain a complete network 
in which all nodes are connected to each other \citep[see][for binary observed variables]{MarsmanEtAL_2018_MBR}. The associated network is shown in Figure \ref{fig:latent-network}, right panel. 
This result may seem counterintuitive, especially if FWMK are correct that partial correlations remove the variance that is shared among variables in the network. In the ULVM, there is, in principle, no unique variance, and all variance can be attributed to a single (latent) variable. We will review this idea later in more detail.   

We only require the following standard assumptions about the latent variable model to obtain our result. 
The random variables $\eta$ and $X$ are such that they satisfy
\begin{itemize}
\item[1.] {\em local independence}: $X_{i}\indep X_{j}\mid \eta$ for all $i\ne j\in V$,
\item[2.] {\em unidimensionality}: $\eta$ is a scalar, and
\item[3.] {\em monotonicity}: if $\eta_{1}>\eta_{2}$ then $\mathbb{P}(X_{j}\mid \eta_{1})>\mathbb{P}(X_{j}\mid \eta_{2})$  for all $j\in V.$
\end{itemize}
Using these assumptions we obtain a marginal distribution of the variables $X_1, X_2, \dots, X_p$ with variance matrix (see the Appendix for a proof)
\begin{align}\label{eq:covaraince-matrix}
\boldsymbol{\Sigma} = \boldsymbol{\lambda}\boldsymbol{\lambda}^{\sf T} + \mathbf{I}_{p}
\end{align}
which is exactly the same as the variance matrix that we observed for the marginal distribution under the ULVM in Section \ref{sec:latent-variable-model}. 
As we saw in Section \ref{sec:gaussian-graphical-model}, a network is obtained by taking the inverse of $\Sigma$, that is $\Sigma^{-1}=\Theta$, which we refer to as the concentration matrix. 

The concentration matrix is (see the Appendix)
\begin{align}\label{eq:inverse-covariance}
\boldsymbol{\Theta}=\mathbf{I}_{p} - \frac{1}{\boldsymbol{\lambda}^{\sf T}\boldsymbol{\lambda}+1}\boldsymbol{\lambda}\boldsymbol{\lambda}^{\sf T}.
\end{align}
We now see that an off-diagonal element $\Theta_{ij}$ for $i\ne j$ is $\alpha\lambda_{i}\lambda_{j}$, where $-\alpha^{-1}=(\boldsymbol{\lambda}^{\sf T}\boldsymbol{\lambda}+1)$. Hence, $\Theta_{ij}$ is in general non-zero. If $\Theta_{ij}=0$, then $\lambda_{i}=0$ for some $i\in V$ and then the variable $X_{i}$ cannot be an indicator variable for the latent variable. Hence, we do not have a ULVM.

We illustrate (\ref{eq:covaraince-matrix}) and (\ref{eq:inverse-covariance}) using $\boldsymbol{\lambda}=(1,0.5,0.5)^{\sf T}$. Then we obtain 
\begin{align*}
\boldsymbol{\Sigma}=
\begin{pmatrix}
2	&0.5		&0.5\\
0.5	&1.25	&0.25\\
0.5	&0.25	&1.25
\end{pmatrix}
\quad \text{and} \quad
\boldsymbol{\Theta}=
\begin{pmatrix}
0.6	&-0.2	&-0.2\\
-0.2	&0.9		&-0.1\\
-0.2	&-0.1	&0.9
\end{pmatrix}
\end{align*}
Computing the element $\Theta_{12}$ using (\ref{eq:inverse-covariance}) with $\boldsymbol{\lambda}^{\sf T}\boldsymbol{\lambda}=1^{2}+0.5^{2}+0.5^{2}=1.5$ gives
\begin{align*}
\alpha\lambda_{1}\lambda_{2} = -\frac{1}{1.5+1}1\cdot 0.5 = -\frac{0.5}{2.5}=-0.2
\end{align*}
which is equivalent to element $\Theta_{12}=-0.2$ in the inverse covariance matrix above. This also shows that for any of the partial covariances $\Theta_{ij}$ to be 0, one of the $\lambda_{i}$ has to be 0. But, obviously, then indicator $i$ is not part of the ULVM.


This result is in line with that of \citet[][Thm 6]{HollandRosenbaum1986}. \citeauthor{HollandRosenbaum1986} showed that a ULVM induces non-zero partial correlations. Suppose that a latent variable model satisfies 1-3 above, then Theorem 6 of \citet{HollandRosenbaum1986} shows that for any partition of the nodes any two nodes are conditionally associated given the other partition. This implies that the partial correlations are all non-zero. \citet{Junker:1997} explain this by saying that the monotone and unidimensional latent variable $\eta$ induces so much `internal coherence' among the observed variables, that the covariation must be larger than 0. These results underscore our concerns with the ideas of FWMK about partial correlation networks. 

Another result, given in \citet{Junker:1997}, shows that when the number of variables that is conditioned on is countably infinite, the covariation vanishes (vanishing conditional independence). This is because an infinite set of highly related variables is an exact (almost sure, in fact) representation of the unidimensional latent variable (or the sigma-field associated with the set of variables conditioned on). In other words, the latent variable $\eta$ can be represented by an {\em infinite} set of variables that are on equal footing with all other variables (i.e., variables that have a similar relation to the latent variable as all others). This result implies that only a network with an infinite number of variables, where all variables would fit the ULVM, will be empty, since in that case the conditioning variables become a representation of the latent variable. This can be seen from the matrix $\boldsymbol{\Theta}$ above, since if there are infinitely many observed variables and $\sum_{i}\lambda_{i}^{2}$ does not converge, then the term $\boldsymbol{\lambda}^{\sf T}\boldsymbol{\lambda}\to \infty$, and so $\boldsymbol{\Theta}$ will tend to $\mathbf{I}_{p}$ as $p$ gets large \citep[see also][equation (7), for a similar result]{Guttman:1953}.

\section{The GGM as a series of regressions}
\label{sec:networks-regression}
A GGM can be estimated by a series of regressions. The reason is that the regression coefficients can be written in terms of the concentration matrix (inverse covariance matrix) of the nodes. Recall that $\Theta_{ij}$ denotes the partial covariance between variables $X_{i}$ and $X_{j}$ with all other variables partialled out, and also recall that if $\Theta_{ij}=0$ this implies that $X_{i}$ and $X_{j}$ are independent conditional on all other variables under consideration. The regression coefficient $\hat{\beta}_{ij}$ can be written in terms of the concentration matrix as \citep[][Section 5.1.3]{Lauritzen96}
\begin{align}
\beta_{ij}=-\frac{\Theta_{ij}}{\Theta_{ii}}
\end{align}
Clearly, if $\Theta_{ij}=0$, then $\beta_{ij}=0$ as well. And so, by inspecting the regression coefficients we can determine the conditional independencies that also hold for the concentration matrix $\Theta$. In the Appendix we provide a small example with three nodes to show that these relations hold. 
Here, we use these relations to show that the regression coefficients indeed explain the dependent variable, which implies that the partial correlations do use the shared variance.

The procedure of using a series of regressions to obtain a GGM was first shown to lead to correct networks in \citet{Meinshausen:2006}.  
We start at any node $i$, and use the associated random variable $X_{i}$ and then call this node $Y$. Then we estimate the non-zero regression coefficients $\beta_{ij}$ for all other remaining nodes in $V\backslash \{i\}$. The notation $\beta_{ij}$ means we are thinking of the connection $i \leftarrow j$ in the network. So, we have a multiple regression, where $Y$ is variable $X_{i}$ and the other variables $X_{V\backslash\{i\}}$ are the predictors
\begin{align}
Y = \beta_{0} + \beta_{i1}X_{1} + \beta_{i2}X_{2} + \cdots + \beta_{ip}X_{p} + e_{i}
\end{align}
where we exclude the predictor $X_{i}$ because we have made that node the dependent variable $Y$.
The non-zero coefficients $\beta_{ij}$ tell us which nodes $j$ are in the neighbourhood of variable $i$, i.e., to which other nodes variable $i$ is connected. 
We do this for all nodes in $V$ and then combine the results because we have used both $\beta_{ij}$ and $\beta_{ji}$, once with variable $j$ being the predictor and once with variable $j$ as the dependent variable. 
We can use the {\em and} rule or the {\em or} rule. 
In the {\em and} rule we say that the edge $i-j$ is present in the network whenever both $\beta_{ij}\ne 0$ and $\beta_{ji}\ne 0$. In the {\em or} rule we identify the edge $i-j$ whenever either $\beta_{ij}\ne 0$ or $\beta_{ji}\ne 0$. 
The idea of estimating the inverse covariance matrix can also be motivated by looking to identify the joint probability distribution of the variables $X_{1},X_{2},\ldots,X_{p}$.
This requires aggregating across all configurations of the random variables, which is computationally difficult. 
One way to make this easier is by reducing the joint distribution into smaller parts and instead of considering all variables simultaneously we only have to consider joint distributions of a smaller number of variables at a time. In the extreme case we use a product of univariate conditional distributions. 
\begin{align}
p(x) \propto p_{1}(x_{1}\mid x_{V\backslash\{1\}})p_{2}(x_{2}\mid x_{V\backslash\{2\}})\cdots p_{p}(x_{p}\mid x_{V\backslash\{p\}})
\end{align}
This is known as the pseudo-likelihood, because it is proportional to the likelihood \citep{Hyvarinen:2006,Nguyen:2017}. 
Each univariate conditional distribution then implies a multivariate regression. To see this, let $Y=X_{i}$ as before and consider the conditional expectation of $Y$ given all remaining variables $X_{V\backslash\{i\}}$
\begin{align}
\mathbb{E}(Y\mid X_{V\backslash \{i\}}) = \beta_{0} + \beta_{i1}X_{1} +\beta_{i2}X_{2} + \cdots + \beta_{ip}X_{p}
\end{align}
This is clearly the regression equation which we consider for each node $i\in V$. Hence by considering all univariate conditional distributions we are in fact determining the pseudo-likelihood which is proportional to the joint density. This idea is related to the coupling of the cliques in the graph and the factorisation of cliques, and is called the Hammerlsey-Clifford theorem \citep[see, e.g.,][]{Lauritzen96,Wainwright:2019}.

\section{Partial correlations and explained variance in regression}
\label{sec:partial-correlation-regression}
Since the GGM coincides with a series of regressions, each node is explained by the remaining nodes in the network. Specifically, at the population level, each node is explained by its neighbors, and the others are irrelevant in the sense that the other variables are independent, given the neighbors. The reason that we can consider the series of regressions the same as a Gaussian graphical model is because of the relation with the conditional covariances, as we saw in the previous part of Section \ref{sec:networks-regression}. 
We first explain the relation between the regression coefficients and the partial correlation more precisely here. Then, we decompose the $R^{2}$ measure and then show with a small example and some simulated data how the explained variance can be (re)distributed among the predictors.

In regression the coefficients are often obtained by ordinary least squares (see the Appendix) and $R^{2}$ is calculated using these coefficients. Suppose we have three variables $X_{1}$, $X_{2}$ and $X_{3}$, as in the example from the introduction corresponding to Figure \ref{fig:explained-variance}. We consider $X_{3}$ as the dependent variable in a regression, so that $X_{1}$ and $X_{2}$ are predictors. If we assume that all three variables have mean 0 and variance 1, then we obtain the regression coefficient (see the Appendix)
\begin{align}\label{eq:beta-3nodes}
\beta_{31}=
\frac{
	\text{cor}(X_{1},X_{3})- 
	\text{cor}(X_{1},X_{2})\text{cor}(Y,X_{2})
	}
	{
	1-\text{cor}(X_{1},X_{2})^{2}
	}
\end{align}
where $\text{cor}()$ is the correlation and $1-\text{cor}(X_{1},X_{2})^{2}$ is the conditional variance of $X_{1}$ given $X_{2}$.  This gives the relation between the regression coefficient and the partial correlation (see the Appendix and \citet{Anderson:1958})
\begin{align}
\beta_{31} \frac{\sqrt{1-\text{cor}(X_{1},X_{2})^{2}}}{\sqrt{1-\text{cor}(X_{3},X_{2})^{2}}} = 
\frac{
	\text{cor}(X_{1},X_{3})- 
	\text{cor}(X_{1},X_{2})\text{cor}(X_{3},X_{2})
	}
	{
	\sqrt{1-\text{cor}(X_{1},X_{2})^{2}}\sqrt{1-\text{cor}(X_{3},X_{2})^{2}}
	}
	=\rho_{31\mid 2}
\end{align}
So the regression coefficient is a rescaling of the partial correlation, where it is clear that both the regression coefficient and the partial correlation use the conditional covariance between $X_{1}$ and $X_{3}$ given $X_{2}$. It is also clear from this formulation that in the coefficient the part of $X_{2}$ is taken out of the correlation between $X_{1}$ and $X_{3}$. 

The fact that the partial covariances are obtained from the covariances by taking its inverse provides the first argument that shows that no information is lost when using the partial covariances (or correlations) to describe the relations between the variables. This is because the covariance and partial covariance are in one-to-one correspondence with each other. That is, for each pair of variables with partial covariance $a$ (point in the space of the partial covariances) there is a unique pair of variables with covariance $b$ (point in the space of covariances). Hence, we can go back and forth from the space of partial covariances and covariances (see the Appendix for a more formal discussion of this). 

The second argument that shows that no information is lost by considering the partial covariances (or partial correlations) comes from considering a GGM as a series of regressions, and the associated multiple correlation measure $R^{2}$ used in regression and in networks \citep{Haslbeck:2018}. The definition of $R^2$ is (see the Appendix)
\begin{align}\label{eq:r2-decomposed}
    R^2=\frac{\text{var}(\hat{Y})}{\text{var}(Y)}=\sum_{i=1}^{p}\beta_{Yi}\frac{\text{cov}(X_{i},Y)}{\text{var}(Y)}
    \end{align}
In other words, we can decompose the explained variance $R^{2}$ into a term for each predictor separately. From this decomposition and (\ref{eq:beta-3nodes}) it is clear that the coefficient represents the unique contribution of the predictor, but that the covariance between the predictor and the dependent variable (not a partial covariance) co-determines the explained variance in regression. 

We consider the three node example of Figure \ref{fig:explained-variance}. Suppose that we take $X_{3}$ as $Y$, the dependent variable, with the predictors $X_{1}$ and $X_{2}$. Then we see that $R^{2}$ is made up of the scaled covariance between each of the predictors and the dependent variable $\text{cov}(X_{i},X_{3})/\text{var}(X_{3})$ multiplied by its respective regression coefficient $\beta_{3i}$. The explained variance part of $X_{1}$ is therefore composed of the complete overlap between $Y$ and $X_{1}$ (scaled by $\text{var}(X_{3})$, c.f. Figure \ref{fig:explained-variance}(b)) and the coefficient $\beta_{31}$ which does not depend on $X_{2}$. The contribution to $R^{2}$ of each predictor is therefore proportional to its covariance (overlap) with the dependent variable. Hence, if we were to take out (partial out) the part of $X_{1}$ out of $X_{2}$ we will not change $R^{2}$ but only redistribute the contribution to $R^{2}$ of each of the predictors. We will show this in simulated dataset, after we briefly discuss the interpretation of the regression coefficients and its relation to partial correlation.

We illustrate the principle of $R^{2}$ and its decomposition in (\ref{eq:r2-decomposed}) further with a small simulation. We generate data according to 
\begin{align*}
X_{3}=\beta_{31}X_{1} +\beta_{32}X_{2} + e
\end{align*}
where we set the coefficients to $\beta_{31}=1$ and $\beta_{32}=2$, respectively, and the error variance to $1$. 
To introduce a correlation (i.e., overlap) between the regressors $X_{1}$ and $X_{2}$, we express $X_2$ in terms of $X_1$ and an additional error term
\begin{align*}
X_{2} = 0.2X_{1} + e_{2}
\end{align*}
where the second error's variance is also set to $1$, so that $\text{cor}(X_{1},X_{2})=0.2$ because $\text{cov}(X_{1},0.2X_{1}+e_{2})=0.2\text{var}(X_{1})+0$ because $\text{var}(X_{1})=1$. We have simulated $n = 100$ cases from this model.

We start with standard regression, which is the default in most statistical packages. The results for the standard regression of $X_3$ on $X_1$ and $X_2$ are shown on the left side of Table \ref{tab:regression-simulations} (the R-syntax for the simulation is in the Appendix). 
From Table \ref{tab:regression-simulations} (left column), we see that the coefficients approximate the population values, and that the predictors explain $87.39\%$ of the response variable. We can now verify the decomposition of (\ref{eq:r2-decomposed}). For this example with three variables we have the decomposition
\begin{align*}
\hat{R}^{2}=\hat{\beta}_{31}\frac{\widehat{\text{cov}}(X_{1},X_{3})}{\widehat{\text{var}}(X_{3})}
	+ \hat{\beta}_{32}\frac{\widehat{\text{cov}}(X_{2},X_{3})}{\widehat{\text{var}}(X_{3})}
\end{align*}
With the values in Table \ref{tab:regression-simulations} and $\widehat{\text{var}}(X_{3})=7.63438$, we obtain
\begin{align*}
\hat{R}^{2}=1.05327\frac{1.62573}{7.63438} + 2.07496\frac{2.39005}{7.63438}=0.22429+0.64960=0.8739
\end{align*}
We see the different contributions of each of the predictors to $R^{2}$, which depends on the combination of the covariance (size of the overlap between predictor and dependent variable) and the regression coefficient. Each regression coefficient has the effect of other variables partialled out, and the contribution of the predictor to $R^{2}$ is determined by the overlap (without anything partialled out) between $X_{1}$ and $X_{3}$ (and scaled by the variance of $X_{3}$ in this example). 

\begin{table}[b]\centering
\caption{Regression output of the small simulation with three random variables.}
\label{tab:regression-simulations}
\begin{tabular}{l		l	l	l @{\hspace{4em}}	l	l	l}
\midrule
		&\multicolumn{3}{c}{standard (type II)}		&\multicolumn{3}{c}{projected (type I)}\\
		&estimate		&std. error		&$\widehat{\text{cov}}(X_{i},X_{3})$		&estimate		&std. error		&$\widehat{\text{cov}}(X_{i},X_{3})$\\
\midrule
$X_{1}$	&1.05327		&0.09482		&1.62573				&1.43143 		&0.09310		&1.62573\\
$X_{2}$	&2.07496    	&0.09863		&2.39005				&2.07496 		&0.09863		&2.09377\\
\midrule
		&\multicolumn{3}{c}{\hspace{-4em}$\widehat{\text{cov}}(X_{1},X_{2})=0.21633$, $\hat{R}^{2}=0.8739$}	&\multicolumn{3}{c}{\hspace{-1em}$\widehat{\text{cov}}(X_{1},X_{2}^{p})=0.01290$, $\hat{R}^{2}=0.8739$}\\
\midrule
\end{tabular}
\end{table}

Next, we do the same but now we first partial out the variance (overlap) of $X_{1}$ from $X_{2}$ before we enter it in the regression. This corresponds to Figure \ref{fig:explained-variance}(b). We consider the regression of $X_3$ on $X_1$ and $X_2^{p}$, where the projected variable ensures that $\text{cor}(X_{1}, X_{2}^{p})=0$. If we believe that $R^{2}$ leaves out completely the overlap between $X_{1}$ and $X_{2}$ (Figure \ref{fig:explained-variance}(c)), then the use of the projected predictor would increase (or remain the same if there were no overlap) the percentage of explained variance, since $X_{1}$ now contains this overlap. This type of regression is sometimes referred to as type I sum of squares, while the former (standard) regression is referred to as type II sum of squares \citep{Ip:2001,Kennedy:2002}. The results for the projected regression are shown on the right of Table \ref{tab:regression-simulations} and reveals a higher coefficient for the first predictor but the same percentage of explained variance. 
The coefficient for $X_{1}$ is higher because we removed any overlap between $X_{1}$ and $X_{2}$ from $X_{2}$, and so we now allow all variance of $X_{1}$ to be explained by $X_{1}$, as in Figure \ref{fig:explained-variance}(b). From (\ref{eq:beta-3nodes}) we clearly see that because $\text{cor}(X_{1},X_{2}^{p})=0$, the coefficient (with the specific settings of means of 0 and variances of 1) is the same as the correlation between $X_{1}$ and $X_{3}$; nothing of $X_{2}^{p}$ is left to subtract from $\text{cor}(X_{1},X_{3})$. 

We verify the decomposition of the explained variance of (\ref{eq:r2-decomposed})
\begin{align*}
\hat{R}^{2}=1.43143\frac{1.62573}{7.63438} + 2.07496\frac{2.09377}{7.63438}=0.30482+0.56910=0.8739
\end{align*}
From this decomposition with $X_{2}^{p}$ instead of $X_{2}$ we notice two things. First, the coefficient $\hat{\beta}_{31}$  increased because the covariance (overlap) between $X_{1}$ and $X_{2}^{p}$ is approximately 0 (see Table \ref{tab:regression-simulations}). From (\ref{eq:beta-3nodes}) this implies that (almost) nothing is ${1}$ and $X_{3}$ because $\widehat{\text{cov}}(X_{1},X_{2}^{p})=0.01290$. So, the coefficient increased from 1.05327 to 1.43143. So, $X_{1}$ is allowed to explain more of the variance of $X_{3}$. The second difference in the $R^{2}$ decomposition is that the covariance $\widehat{\text{cov}}(X_{2}^{p},X_{3})$ is reduced from 2.39005 to 2.09377 because the common part with $X_{1}$ is taken out of $X_{2}$ giving the variable $X_{2}^{p}$. These two changes lead to different decompositions in $R^{2}$. But, obviously, owe have not changed to total variance (area) of $X_{3}$ explained by the predictors $X_{1}$ and either $X_{2}$ or $X_{2}^{p}$. The only thing that has changed is which predictor gets to explain the variance of $X_{3}$. 

Since in the projected regression we took out of $X_{2}$ anything that was in common with $X_{1}$, and $R^{2}$ is exactly the same, we must conclude that a standard regression indeed explains all the variance that can be explained by the predictors. That is, no shared variance is taken out.


\section{Regularised regression and the GGM}
Although the relation between regression, GGM and networks is clear from the previous sections, in practice, regression is often performed with some alternative way that may change the relation with the original network. Here we will focus on the least absolute shrinkage and selection operator (lasso, or $\ell_{1}$-norm) in regression. This regularisation technique takes the sum of the absolute values of the parameters $\beta_{ij}$ as a penalty, i.e., $\sum_{j=1}^{p}|\beta_{ij}|$. Because this function is also minimised, the lasso shrinks the parameter values towards zero, or sets them to zero, depending on the regularisation parameter \citep{Tibshirani:1996}. It has been shown that, given a set of assumptions, the lasso is consistent, meaning that the correct parameters are obtained in a specific asymptotic framework \citep[e.g.,][]{Meinshausen:2006,Wainwright:2009,Buhlmann:2011,Waldorp:2019}. One of the assumptions of the lasso is that the network is sparse, i.e., in the situation of a network where nodes are added at each step, the number of edges will always remain bounded (the number of edges is in the order of the number of nodes). For a dense network, however, the parameters will be poorly estimated \citep{Waldorp:2019}. As a consequence for dense networks, the regression parameters of the lasso are inappropriate to use as scaled partial correlations because many of the edges will have been set to 0, while they should be part of the network. In the extreme case discussed in this manuscript, the ULVM corresponds to a fully-connected network, and so, the lasso will incorrectly set several edges to 0. Although this does not change the results of the previous sections, it does warrant careful consideration if the network to be estimated is sparse or dense.

\section{Concluding Comments}\noindent
In this paper, we have refuted the belief that partial correlations remove the shared variance in estimating GGMs, as recently voiced by FWMK, and have shown that all variance of the focal node that can be explained by other nodes is explained. First, we showed that if the data come from a ULVM, and there is no unique variance, the estimated network is fully-connected, and not empty, as FWMK would make us believe. Secondly, we have revisited the relation between GGMs, partial correlations, and regression to show that partial correlations indeed do not remove shared variance from the explained variance.

We have also established a formal connection between the latent variable model and the GGM, which is further evidence for broad connections that exist between graphical models and latent variable models. A particular consequence of these relations is that reliability and replication issues for one model, an unrestricted GGM, say, are also likely to be an issue for the other. It is interesting to observe that the critique of FWMK has focused on one of the two models while advocating the other, which seems contradictory given these formal results. 

This incongruity leaves us with what we believe is the most relevant issue, not mentioned by FWMK, but certainly present between the lines: The network model is wrong. The network model may indeed be wrong, and this is worth discussing and investigating scientifically. We believe that one of the most important ways to approach such a debate is by considering what predictions a model makes and how this can be verified or falsified empirically. 

\section*{Appendix}\noindent
Here we provide the mathematical details and proofs for the claims that have been made in the main text. 
We provide these details in the same order as in the main text. 
First, we formalise the ideas about the marginal density of variables under the ULVM. Then, we provide some details regarding the relation between GGMs and regression. And finally, we provide the details about the projection matrices.

\subsection*{The Marginal Distribution under the ULVM}
We show that the marginal of the observed variables over the latent variable is a fully connected graph. We use the assumptions and its consequences for the marginal distribution over the observed variables $X_{1}, X_{2}, \ldots, X_{p}$ given in the main text. This has already been shown for binary data modeled by the Ising model \citep{MarsmanEtAL_2018_MBR}. Here we show a similar result for Gaussian data. We do this by assuming that what we observe is the marginal distribution $p(\mathbf{x})=\int p(\mathbf{x}\mid \eta)p(\eta)d\eta$ which results in all observed variables in $\mathbf{x}$ being connected. 

We assume that there is a unidimensional latent variable $\eta$ that is Gaussian distributed with mean 0 and variance 1 (for convenience). This is Assumption 1 in Section \ref{sec:relations-networks-latent}. Assumption 2 states that conditional on $\eta$ the $p$ random variables in $\mathbf{x}$ are multivariate normal with mean $\boldsymbol{\lambda} \eta$, where the vector $\boldsymbol{\lambda}$ are the loadings, and variance matrix $\mathbf{I}_{p}$ (isotropic variance matrix). Assumptions 3 from Section \ref{sec:relations-networks-latent} (monotonicity) is not strictly necessary for this derivation, but according to \cite{Junker:1997} monotonicity is required in order for the model to be `useful', which we agree with. These assumptions together imply that the conditional density of the observed values $x_{j}$, $j = 1, 2, \dots, p$ is equal to
\begin{align*}
p(\mathbf{x}\mid \eta) = \frac{1}{\sqrt{(2\pi)^{p}}} \exp\left( -\tfrac{1}{2}(\mathbf{x}-\boldsymbol{\lambda} \eta)^{\sf T}(\mathbf{x}-\boldsymbol{\lambda} \eta)\right)
\end{align*}
and that the prior distribution for the latent variable is equal to 
\begin{align*}
p(\eta)=(2\pi)^{-1/2}\exp(-\tfrac{1}{2}\eta^{2})
\end{align*}
The joint distribution $p(\mathbf{x},\eta)=p(\mathbf{x}\mid \eta)p(\eta)$ is then
\begin{align*}
p(\mathbf{x}, \eta) = \frac{1}{\sqrt{(2\pi)^{p+1}}} \exp\left( -\tfrac{1}{2}(\mathbf{x}-\boldsymbol{\lambda} \eta)^{\sf T}(\mathbf{x}-\boldsymbol{\lambda} \eta)-\tfrac{1}{2}\eta^{2}\right)
\end{align*}

We use the unidimensional Gaussian integral $\int \exp(-ax^{2}+bx +c)dx=(2\pi)\exp(b^{2}/4a +c)$. The marginal now becomes
\begin{align*}
\int_{\mathbb{R}} \exp\left( -\tfrac{1}{2}(\mathbf{x}-\boldsymbol{\lambda} \eta)^{\sf T}(\mathbf{x}-\boldsymbol{\lambda}\eta) - \tfrac{1}{2}\eta^{2}\right) d\eta
=
\sqrt{(2\pi)^{p}}\exp\left( -\tfrac{1}{2}\mathbf{x}^{\sf T}(\boldsymbol{\lambda}\boldsymbol{\lambda}^{\sf T}+\mathbf{I}_{p})^{-1} \mathbf{x} \right)
\end{align*}
The variance matrix $\boldsymbol{\Sigma}=\boldsymbol{\lambda}\boldsymbol{\lambda}^{\sf T}+\mathbf{I}_{p}$ is positive definite since we have the square term $\boldsymbol{\lambda}\boldsymbol{\lambda}^{\sf T}$. The inverse of the variance matrix $\boldsymbol{\Sigma}$ is
\begin{align*}
(\boldsymbol{\lambda}\boldsymbol{\lambda}^{\sf T}+\mathbf{I}_{p})^{-1} = \mathbf{I}_{p} - \frac{1}{\boldsymbol{\lambda}^{\sf T}\boldsymbol{\lambda}+1}\boldsymbol{\lambda}\boldsymbol{\lambda}^{\sf T}
\end{align*}

which can be obtained by the Sherman-Morrison theorem \citep[e.g.,][]{BilodeauBrenner99}. 

It is easy to check that this is the inverse
\begin{align*}
(\boldsymbol{\lambda}\boldsymbol{\lambda}^{\sf T}+\mathbf{I}_{p})^{-1}(\boldsymbol{\lambda}\boldsymbol{\lambda}^{\sf T}+\mathbf{I}_{p}) = 
	\mathbf{I}_{p} 
	+\boldsymbol{\lambda}\boldsymbol{\lambda}^{\sf T}
	-  \frac{1}{\boldsymbol{\lambda}^{\sf T}\boldsymbol{\lambda}+1}\boldsymbol{\lambda}\boldsymbol{\lambda}^{\sf T}
	- \frac{\boldsymbol{\lambda}^{\sf T}\boldsymbol{\lambda}}{\boldsymbol{\lambda}^{\sf T}\boldsymbol{\lambda}+1}\boldsymbol{\lambda}\boldsymbol{\lambda}^{\sf T}
\end{align*}
and taking the last two elements together yields
\begin{align*}
\mathbf{I}_{p} +  \boldsymbol{\lambda}\boldsymbol{\lambda}^{\sf T}
		- \frac{\boldsymbol{\lambda}^{\sf T}\boldsymbol{\lambda}+1}{\boldsymbol{\lambda}^{\sf T}\boldsymbol{\lambda}+1}
		 \boldsymbol{\lambda}\boldsymbol{\lambda}^{\sf T}
	=\mathbf{I}_{p}
\end{align*}

We immediately see that the observed variables are all correlated if all of the connections from the latent variable $\eta$ to the observed values $\mathbf{x}$ in $\boldsymbol{\lambda}$ were non-zero. In some special cases (with measure 0) values in $(\boldsymbol{\lambda}\boldsymbol{\lambda}^{\sf T}+\mathbf{I}_{p})^{-1}$ could cancel and become zero \citep[e.g.,][Chapter 13]{Buhlmann:2011}. It can also be seen that whenever the number of variables increases to infinity, then the partial covariances (and hence correlations) will become 0. 
Of course, if this sum converges to $\gamma$, say, then we find that the partial covariance is $1-(\gamma+1)^{-1}\lambda_{i}\lambda_{j}$ for the variables $i$ and $j$.  For this to happen, infinitely many $\lambda_{i}^{2}$ will have to be arbitrarily close to 0. This implies that the relation between the latent variable and the indicator is 0, and so it cannot be a ULVM.

\subsection*{Obtaining a GGM by Regressions}
We can estimate the coefficients $\boldsymbol{\beta}_{i}=(\beta_{i1},\beta_{i2},\ldots,\beta_{ip})$, without $\beta_{ii}$, of course, by using least squares. We often omit the index $i$ for node $i\in V$ because the procedure is generic for all $i$ in $V$. Then we minimise the squared residuals (ordinary least squares)
\begin{align*}
||\mathbf{y}-\mathbf{X}\boldsymbol{\beta}||^{2}=\sum_{k=1}^{n}(y_{k}-\mathbf{x}_{k}\beta)^{2}
\end{align*}
where $\mathbf{x}_{k}$ denotes the vector of $p-1$  predictors of nodes $V\backslash \{i\}$ and the intercept (constant) for observation $k$, i.e., row $k$ of the $n\times p$ matrix $\mathbf{X}$. We then obtain the well known ordinary least squares estimate \citep[see, e.g., ][Chapter 5]{BilodeauBrenner99}
\begin{align*}
\hat{\boldsymbol{\beta}} = (\mathbf{X}^{\sf T}\mathbf{X})^{-1}\mathbf{X}^{\sf T}\mathbf{y}
\end{align*}
We subtract the main effect of the predictors in $\mathbf{X}$ and $\mathbf{y}$, but keep the same names, so that their means are 0. We can then immediately see that this estimate $\hat{\boldsymbol{\beta}}$ can be rewritten as
\begin{align*}
\hat{\boldsymbol{\beta}} = \text{var}(\mathbf{X})^{-1}\text{cov}(\mathbf{X},\mathbf{y})
\end{align*}
where $\text{var}(\mathbf{X})$ is the $p\times p$ variance matrix of the predictors in $\mathbf{X}$ and $\text{cov}(\mathbf{X},\mathbf{y})$ is the $p\times 1$ vector of covariances between $\mathbf{y}$ and each $\mathbf{x}_{j}$ for $j\in V\backslash \{i\}$. It is convenient to think of the least squares estimate in this way because we will couple the estimate to Gaussian graphical models later on. 

We assume that we modeled correctly, so that the residuals $e_{i}$ are uncorrelated and have the unit variance across observations, i.e., 
\begin{align*}
\mathbb{E}(\mathbf{y}\mid \mathbf{X}_{V\backslash \{i\}})= \beta_{0} + \beta_{i1}\mathbf{x}_{1} + \beta_{i2}\mathbf{x}_{2} + \cdots + \beta_{ip}\mathbf{x}_{p} 
\end{align*}
In a GGM we are interested in $\boldsymbol{\Theta}=\boldsymbol{\Sigma}^{-1}$, because if we find $\Theta_{ij}=0$ then we have that $X_{i}\indepen X_{j}\mid X_{V\backslash \{i,j\}}$ \citep{Lauritzen96}. In fact, we have that $\beta_{ij}=-\Theta_{ij}/\Theta_{ii}$. And so, if $\beta_{ij}=0$ then obviously, we must have that $\Theta_{ij}=0$, and vice versa. So, by inspecting the regression coefficients, we are in fact considering the concentration matrix.

For the random variables$X_{i}$ and $Y$ and the predicted value $\hat{Y}=\sum_{i=1}^{p}X_{i}\beta_{Yi}$, the squared multiple correlation $R^{2}$ can be defined by 
\begin{align*}
R^{2} = \text{cor}(\hat{Y},Y)^{2}=\frac{\text{var}(\hat{Y})}{\text{var}(Y)}
\end{align*}
The variance of the predicted values can be written as $\text{var}(\hat{Y})=\text{cov}(\sum_{i=1}^{p}X_{i}\beta_{Yi},Y)$. And this gives the decomposition
\begin{align*}
R^{2} = \sum_{i=1}^{p}\beta_{Yi}\frac{\text{cov}(X_{i},Y)}{\text{var}(Y)}
\end{align*}
This and other decompositions are given in e.g., \citet{Genizi:1993}.

\subsection*{Relation between partial correlation and regression}
Suppose we have three variables $X_{1}$, $X_{2}$ and $X_{3}$. We consider $X_{3}$ the dependent variable, so that $X_{1}$ and $X_{2}$ are predictors, and we are mostly interested in the situation where we partial out $X_{2}$. If we assume that all three variables have 0 mean and variance 1, then we have that the conditional variances are
\begin{align*}
\text{var}(X_{1}\mid X_{2}) &= \text{var}(X_{1})-\text{cor}(X_{1},X_{2})^{2}=1-\text{cor}(X_{1},X_{2})^{2}\quad \text{and}\\
\text{var}(X_{3}\mid X_{2}) &=\text{var}(X_{1})-\text{cor}(X_{3},X_{2})^{2}=1-\text{cor}(X_{3},X_{2})^{2}
\end{align*}
And the conditional covariance between $X_{1}$ and $X_{3}$ given $X_{2}$ is
\begin{align*}
\text{cov}(X_{1},X_{3}\mid X_{2})=
\text{cor}(X_{1},X_{3})- 
\text{cor}(X_{1},X_{2})\text{cor}(X_{3},X_{2})
\end{align*}
 We can now write the partial correlation in the well-known version
\begin{align*}
\rho_{31\mid 2} 
&=
\frac{
	\text{cov}(X_{1},X_{3}\mid X_{2})
	}
	{
	\sqrt{\text{var}(X_{1}\mid X_{2})} \sqrt{\text{var}(X_{3}\mid X_{2}) }
}\\
&=
\frac{
	\text{cor}(X_{1},X_{3})- 
	\text{cor}(X_{1},X_{2})\text{cor}(X_{3},X_{2})
	}
	{
	\sqrt{1-\text{cor}(X_{1},X_{2})^{2}}\sqrt{1-\text{cor}(X_{3},X_{2})^{2}}
	}
\end{align*}
and the regression coefficient $\beta_{13}$ as 
\begin{align*}
\beta_{31}=
\frac{
	\text{cor}(X_{1},X_{3})- 
	\text{cor}(X_{1},X_{2})\text{cor}(X_{3},X_{2})
	}
	{
	\sqrt{1-\text{cor}(X_{1},X_{2})^{2}}
	}
\end{align*}
It is clear that the regression coefficient and the partial correlation use the same information from the conditional covariance $\text{cov}(X_{1},X_{3}\mid X_{2})$ but that the scaling in the partial correlation is also with respect to the conditional variance $\text{var}(Y\mid X_{2})$. 

\subsection*{Proof that $\Sigma^{-1}=\Theta$ is a one-to-one correspondence with $\Sigma$}

We discuss here that a random variable in $\mathbb{R}^{p}$ can be mapped from the space associated with partial covariances to the space associated with covariances. Suppose we have a standard normal random variable $\mathbf{z}\in \mathbb{R}^{p}$, i.e., it has mean $\mathbf{0}$ and variance matrix $\mathbf{I}_{p}$. For the the positive definite variance matrix $\boldsymbol{\Sigma}$ take its symmetric square root $\boldsymbol{\Sigma}^{1/2}$ (e.g., $\boldsymbol{\Sigma}^{1/2}=\mathbf{U}\boldsymbol{\Lambda}^{1/2}\mathbf{U}^{\sf T}$, where $\boldsymbol{\Lambda}^{1/2}=\text{diag}(\lambda^{1/2}_{1},\ldots,\lambda_{p}^{1/2})$ is the diagonal matrix with the square roots of the eigenvalues and $\mathbf{U}$ contains the $p$ eigenvectors). Since $\boldsymbol{\Sigma}$ is positive definite, all eigenvalues are $>0$ and the inverse exists. And so the inverse of $\boldsymbol{\Sigma}^{1/2}$ also exists. Let $\mathbf{x}=\boldsymbol{\Sigma}^{-1/2}z$ such that $\text{var}(\mathbf{x})=\boldsymbol{\Sigma}^{-1}$ are the partial covariances. Then $\mathbf{y}=\boldsymbol{\Sigma} \mathbf{x}$ is a normal random variable with variance matrix $\text{var}(\mathbf{y})=\boldsymbol{\Sigma}$. Hence, if $\boldsymbol{\Sigma}$ is a one-to-one correspondence for variables $\mathbf{x}$ associated with the partial covariances and $\mathbf{y}$ associated with the covariances, then there is a unique relation between any such points in those spaces and the two spaces are basically the same (an isomorphism). We make this more precise in the following.  
 
Let $V,W\subseteq \mathbb{R}^{p}$ be two linear subspaces. Then there is a one-to-one correspondence (isomorphism) between $V$ and $W$ obtained by the symmetric, positive definite linear transformation $\boldsymbol{\Sigma}$ (a bijection). This implies that for each element $\mathbf{v}\in V$ there is a $\mathbf{w}\in W$, and you can go back and forth between them using $\mathbf{w}=\boldsymbol{\Sigma} \mathbf{v}$ and $\mathbf{v}=\boldsymbol{\Sigma}^{-1}w$. In the framework of networks, the matrix $\boldsymbol{\Sigma}$ is identified with the covariances and $\boldsymbol{\Sigma}^{-1}=\boldsymbol{\Theta}$ is identified with the partial covariances. But because there is an isomorphism between the spaces $V$ and $W$ obtained with the bijection $\boldsymbol{\Sigma}$, the spaces $V$ and $W$ can be considered the same (upon relabelling points). A characterisation of $V$ and $W$ being the same (are isomorphic) is that they have the same dimension (i.e., same number of basis vectors). We associate $V$ with the partial covariances and $W$ with the covariances. In the example of the previous paragraph we may write $\mathbf{x}=\boldsymbol{\Sigma}^{-1/2}\mathbf{z}\in V$ and $\mathbf{y}=\boldsymbol{\Sigma} \mathbf{x}\in W$. If $\boldsymbol{\Sigma}$ is a bijection that respects the structure (see below), then the spaces $V$ and $W$ are isomorphic; informally, the spaces are the same and contain the same information.

We show that having an inverse $\boldsymbol{\Sigma}^{-1}=\boldsymbol{\Theta}$ (left and right inverse) implies that $\boldsymbol{\Sigma}$ is a bijection between $V$ and $W$ and vice versa, and for this linear transformation the structure is preserved, i.e., $\boldsymbol{\Sigma}(\mathbf{u}+\mathbf{v})=\boldsymbol{\Sigma} \mathbf{u}+\boldsymbol{\Sigma} \mathbf{v}$ and $\boldsymbol{\Sigma} (\alpha \mathbf{v}) = \alpha (\boldsymbol{\Sigma} \mathbf{v})$ for some $\alpha\in \mathbb{R}$. 

Suppose $\boldsymbol{\Sigma}$ is a bijection. Then it is injective, i.e., if $\boldsymbol{\Sigma} \mathbf{v} = \boldsymbol{\Sigma} \mathbf{v}'$ then $\mathbf{v}=\mathbf{v}'$, and it is surjective, i.e., for each $\mathbf{w}\in W$ there is a $\mathbf{v}\in V$ such that $\boldsymbol{\Sigma} \mathbf{v}=\mathbf{w}$. We define a function $\boldsymbol{\Theta}$ such that for any $\mathbf{v}\in V$ with $\boldsymbol{\Sigma} \mathbf{v}=\mathbf{w}$ we obtain $\boldsymbol{\Theta} \mathbf{w}=\mathbf{v}$ (left inverse). Note that since $\boldsymbol{\Sigma}$ is injective, there is only one such $\mathbf{v}$. We also define $\boldsymbol{\Theta}'$ such that for any $\mathbf{v}\in V$ with $\mathbf{v}^{\sf T}\boldsymbol{\Sigma}=\mathbf{w}^{\sf T}$ we obtain $\mathbf{w}^{\sf T} \boldsymbol{\Theta}' = \mathbf{v}^{\sf T}$ (right inverse). We immediately find that $\boldsymbol{\Theta}=\boldsymbol{\Theta}'$ because
\begin{align*}
\boldsymbol{\Theta} \mathbf{w} = \boldsymbol{\Theta} (\boldsymbol{\Sigma} \mathbf{v}) = \boldsymbol{\Theta} (\boldsymbol{\Sigma} (\mathbf{w}^{\sf T} \boldsymbol{\Theta}')^{\sf T} )= (\boldsymbol{\Theta} \boldsymbol{\Sigma}) (\mathbf{w}^{\sf T} \boldsymbol{\Theta}')^{\sf T} =\boldsymbol{\Theta}' \mathbf{w}
\end{align*}

Now suppose we have a unique $\boldsymbol{\Theta}$ such that $\boldsymbol{\Theta} \mathbf{w}=(\mathbf{w}^{\sf T} \boldsymbol{\Theta})^{\sf T} =\mathbf{v}$ (left and right inverse). Then we have that $\boldsymbol{\Theta} \mathbf{w} =\boldsymbol{\Theta} \mathbf{w}'$ implies $\mathbf{w}=\boldsymbol{\Sigma} (\boldsymbol{\Theta} \mathbf{w} ) = \boldsymbol{\Sigma} (\boldsymbol{\Theta} \mathbf{w}') =\mathbf{w}'$, showing that $\boldsymbol{\Sigma}$ is injective. To show that $\boldsymbol{\Sigma}$ is surjective, choose any $\mathbf{w}\in W$ with $\boldsymbol{\Theta} \mathbf{w} = \mathbf{v}$ for some $\mathbf{v}\in V$. Then $\boldsymbol{\Sigma} \mathbf{v} =\boldsymbol{\Sigma} (\boldsymbol{\Theta} \mathbf{w}) = \mathbf{w}$ showing that $\boldsymbol{\Sigma}$ is surjective. 

The fact that the structure under the mapping $\boldsymbol{\Sigma}$ is preserved (homomorphism) is a direct consequence of linear transformations. This completes the proof.

\subsection*{Projected Predictors and $R^{2}$}
The value used for regression $R^{2}$ is defined as 
\begin{align}
R^{2} = 1 -\frac{ \mathbf{y}^{\sf T}\mathbf{Q}_{X}\mathbf{y}}{\mathbf{y}^{\sf T}\mathbf{y}}
\end{align}
where $\mathbf{Q}_{X}=\mathbf{I}_{n}-\mathbf{P}_{X}=\mathbf{X}(\mathbf{X}^{\sf T}\mathbf{X})^{-1}\mathbf{X}^{\sf T}$ is a projection matrix \citep[see, e.g.,][]{BilodeauBrenner99,Schott97}. A projection matrix has the property that $\mathbf{P}_{X}\mathbf{P}_{X}=\mathbf{P}_{X}$ (idempotent) and $\mathbf{P}_{X}^{\sf T}=\mathbf{P}_{X}$ (symmetric). It can be verified that $\mathbf{Q}_{X}=\mathbf{I}_{n}-\mathbf{P}_{X}$ is also a projection matrix and is orthogonal to $\mathbf{P}_{X}$, i.e., $\mathbf{P}_{X}\mathbf{Q}_{X}=\mathbf{0}$. 

The procedure introduced in Section \ref{sec:partial-correlation-regression} has the predictors ordered so that we start with the first predictor $\mathbf{x}_{1}$ and leave that intact. Then we insert a new predictor $\mathbf{x}_{2}^{p}=\mathbf{Q}_{2}\mathbf{x}_{2}$ such that $\text{cor}(\mathbf{x}_{1},\mathbf{x}_{2}^{p})=0$, where $\mathbf{Q}_{2}=\mathbf{I}_{n}-\mathbf{x}_{1}(\mathbf{x}_{1}^{\sf T}\mathbf{x}_{1})^{-1}\mathbf{x}_{1}^{\sf T}$ and $\mathbf{Q}_{1}=\mathbf{I}_{n}$. We continue including new predictors $\mathbf{x}_{j}$ such that for each pair $\text{cor}(\mathbf{x}_{I},\mathbf{x}_{j})=0$, for all predictors $\mathbf{x}_{i}$ with $i\in I$ and $i<j$ in terms of entering the regression. This can be defined recursively by $\mathbf{Q}_{i+1}=\mathbf{Q}_{i}\mathbf{Q}_{i-1}\cdots \mathbf{Q}_{1}$. This procedure corresponds to the type I sum of squares \citep{Ip:2001,Kennedy:2002}. Figure \ref{fig:explained-variance}(b) shows that the area of $\mathbf{x}_{1}$ remains as is, and that all of $\mathbf{x}_{1}$ is taken out of $\mathbf{x}_{2}$ by $\mathbf{Q}_{2}\mathbf{x}_{2}$. 
This can be achieved in general by
\begin{align*}
(\mathbf{Q}_{1},\mathbf{Q}_{2},\ldots,\mathbf{Q}_{p})
\text{diag}(\mathbf{x}_{1}, \mathbf{x}_{2}, \ldots, \mathbf{x}_{p}) = \mathbf{Q}d(\mathbf{X})
\end{align*}
where $\text{diag}(\mathbf{x}_{1},\ldots,\mathbf{x}_{p})=d(\mathbf{X})$ is the $np\times p$ matrix with $\mathbf{x}_{i}$ on the diagonal and $\mathbf{Q}$ is the $n\times np$ matrix with the orthogonal projects $\mathbf{Q}_{i}$ described above. With this notation we can write the error as $\mathbf{Q}_{X}\mathbf{y}=\mathbf{Q}_{X}\mathbf{e}$, where $\mathbf{e}$ is the residual and $\mathbf{Q}_{X}=\mathbf{I}_{n}-\mathbf{Q}d(\mathbf{X})[d(\mathbf{X})^{\sf T}\mathbf{Q}^{\sf T}\mathbf{Q}d(\mathbf{X})]^{-1}d(\mathbf{X})^{\sf T}\mathbf{Q}^{\sf T}$ is the $n\times n$ projection matrix orthogonal to $\mathbf{X}$. We then have that 
\begin{align*}
\mathbb{E}(\mathbf{y}^{\sf T}\mathbf{Q}_{X}\mathbf{y}) = \mathbb{E}(\mathbf{e}^{\sf T}\mathbf{Q}_{X}\mathbf{e}) =\tr (\mathbf{Q}_{X}\mathbb{E}(\mathbf{e} \mathbf{e}^{\sf T}))=n-p
\end{align*}
because we assumed that $\mathbb{E}(\mathbf{e} \mathbf{e}^{\sf T})=\mathbf{I}_{n}$. So for any reasonable projection such that the rank of the predictors does not change (i.e., $\mathbf{X}^{\sf T}\mathbf{X}$ must be non-singular, no predictors can be correlated too highly), we obtain that the value for $R^{2}$ remains the same for the projected and non-projected regressions.

\subsection*{R Syntax for the simulated data}
{\small
\begin{verbatim}
n <- 100
beta1 <- 1
beta2 <- 2
sigma2 <- 1

set.seed(34)
noise <- rnorm(n)

x1 <- rnorm(n,sd=sigma2)
x2 <- 0.2*x1 + rnorm(n,sd=sigma2)
y <- beta1*x1 + beta2*x2 + noise

fit <- lm(y ~ -1 + x1 + x2) # 1.05327*x1 + 2.07496*x2
summary(fit) # R2 is 0.8738

(1.05327*cov(x1,y) + 2.07496*cov(x2,y))/var(y) # R2 decomposed 0.8738892



X <- cbind(x1)
Xc.proj <- diag(1,n) - X%*%solve(t(X)%*%X)%*%t(X)
x2p <- Xc.proj%*%x2

cor(x1,x2) # 0.201
cor(x1,x2p) # 0.012
cor(cbind(y,x1,x2))
inv.cor <- solve(cor(cbind(x1,x2)))
sqrt(diag(diag(inv.cor)))%*%(-inv.cor)%*%sqrt(diag(diag(inv.cor)))

fitp <- lm(y ~ -1+ x1 + x2p) # 1.43143*x1 + 2.07496*x2
summary(fitp) # R2 is 0.8738

(1.43143*cov(x1,y) + 2.07496*cov(x2p,y))/var(y) # R2 decomposed 0.8738905
\end{verbatim}}
%


\begin{thebibliography}{27}

\bibitem[\protect\citeauthoryear{Anderson}{1958}]{Anderson:1958}
\begin{bbook}[author]
\bauthor{\bsnm{Anderson},~\bfnm{T.~W.}\binits{T.~W.}}
(\byear{1958}).
\btitle{An introduction to multivariate statistical analysis}.
\bpublisher{John Wiley and Sons}, \baddress{New York}.
\end{bbook}
\endbibitem

\bibitem[\protect\citeauthoryear{Bilodeau and
  Brenner}{1999}]{BilodeauBrenner99}
\begin{bbook}[author]
\bauthor{\bsnm{Bilodeau},~\bfnm{M.}\binits{M.}} \AND
  \bauthor{\bsnm{Brenner},~\bfnm{D.}\binits{D.}}
(\byear{1999}).
\btitle{Theory of multivariate statistics}.
\bpublisher{Springer-Verlag}, \baddress{New York}.
\end{bbook}
\endbibitem

\bibitem[\protect\citeauthoryear{B{\"u}hlmann and van~de
  Geer}{2011}]{Buhlmann:2011}
\begin{bbook}[author]
\bauthor{\bsnm{B{\"u}hlmann},~\bfnm{P.}\binits{P.}} \AND
  \bauthor{\bparticle{van~de} \bsnm{Geer},~\bfnm{S.}\binits{S.}}
(\byear{2011}).
\btitle{Statistics for High-Dimensional Data: Methods, Theory and
  Applications}.
\bpublisher{Springer}, \baddress{Berlin}.
\end{bbook}
\endbibitem

\bibitem[\protect\citeauthoryear{Dawid}{1979}]{Dawid1979}
\begin{barticle}[author]
\bauthor{\bsnm{Dawid},~\bfnm{{A. P. }}\binits{A.}}
(\byear{1979}).
\btitle{Conditional Independence in Statistical Theory}.
\bjournal{Journal of the Royal Statistical Society. Series B (Methodological)}
\bvolume{41}
\bpages{1--31}.
\end{barticle}
\endbibitem

\bibitem[\protect\citeauthoryear{Epskamp et~al.}{2018}]{EpskampEtAl_2018_HoP}
\begin{binbook}[author]
\bauthor{\bsnm{Epskamp},~\bfnm{S.}\binits{S.}},
  \bauthor{\bsnm{Maris},~\bfnm{{G. K. J. }}\binits{G.}},
  \bauthor{\bsnm{Waldorp},~\bfnm{{L. J. }}\binits{L.}} \AND
  \bauthor{\bsnm{Borsboom},~\bfnm{D.}\binits{D.}}
(\byear{2018}).
\btitle{Network Psychometrics}
In \bbooktitle{{H}andbook of psychometrics}
\bpages{953--986}.
\bpublisher{Wiley-Blackwell}, \baddress{New York, NY}.
\end{binbook}
\endbibitem

\bibitem[\protect\citeauthoryear{Forbes et~al.}{2017}]{ForbesEtAl_2017}
\begin{barticle}[author]
\bauthor{\bsnm{Forbes},~\bfnm{M.~K.}\binits{M.~K.}},
  \bauthor{\bsnm{Wright},~\bfnm{A.~G.~C.}\binits{A.~G.~C.}},
  \bauthor{\bsnm{Markon},~\bfnm{K.~E.}\binits{K.~E.}} \AND
  \bauthor{\bsnm{Krueger},~\bfnm{R.~F.}\binits{R.~F.}}
(\byear{2017}).
\btitle{Evidence that psychopathology symptom networks have limited
  replicability}.
\bjournal{Journal of Abnormal Psychology}
\bvolume{126}
\bpages{969--988}.
\bdoi{10.1037/abn0000276}
\end{barticle}
\endbibitem

\bibitem[\protect\citeauthoryear{Forbes et~al.}{2019a}]{ForbesEtAl_IP_MBR}
\begin{barticle}[author]
\bauthor{\bsnm{Forbes},~\bfnm{M.~K.}\binits{M.~K.}},
  \bauthor{\bsnm{Wright},~\bfnm{A.~G.~C.}\binits{A.~G.~C.}},
  \bauthor{\bsnm{Markon},~\bfnm{K.~E.}\binits{K.~E.}} \AND
  \bauthor{\bsnm{Krueger},~\bfnm{R.~F.}\binits{R.~F.}}
(\byear{2019}a).
\btitle{Quantifying the reliability and replicability of psychopathology
  network characteristics}.
\bjournal{Multivariate Behavioral Research}.
\bdoi{10.1080/00273171.2019.1616526}
\end{barticle}
\endbibitem

\bibitem[\protect\citeauthoryear{Forbes
  et~al.}{2019b}]{ForbesEtAl_2019_WorldPsychiatry}
\begin{barticle}[author]
\bauthor{\bsnm{Forbes},~\bfnm{M.~K.}\binits{M.~K.}},
  \bauthor{\bsnm{Wright},~\bfnm{A.~G.~C.}\binits{A.~G.~C.}},
  \bauthor{\bsnm{Markon},~\bfnm{K.~E.}\binits{K.~E.}} \AND
  \bauthor{\bsnm{Krueger},~\bfnm{R.~F.}\binits{R.~F.}}
(\byear{2019}b).
\btitle{The network approach to psychopathology: {P}romise versus reality}.
\bjournal{World Psychiatry}
\bvolume{18}
\bpages{272-273}.
\bdoi{10.1002/wps.20659}
\end{barticle}
\endbibitem

\bibitem[\protect\citeauthoryear{Genizi}{1993}]{Genizi:1993}
\begin{barticle}[author]
\bauthor{\bsnm{Genizi},~\bfnm{A.}\binits{A.}}
(\byear{1993}).
\btitle{Decomposition of $R^2$ in multiple regression with correlated
  regressors}.
\bjournal{Statistica Sinica}
\bvolume{3}
\bpages{407-420}.
\end{barticle}
\endbibitem

\bibitem[\protect\citeauthoryear{Guttman}{1953}]{Guttman:1953}
\begin{barticle}[author]
\bauthor{\bsnm{Guttman},~\bfnm{L.}\binits{L.}}
(\byear{1953}).
\btitle{Image theory for the structure of quantitative variates}.
\bjournal{Psychometrika}
\bvolume{18}
\bpages{277--296}.
\end{barticle}
\endbibitem

\bibitem[\protect\citeauthoryear{Haslbeck and Waldorp}{2018}]{Haslbeck:2018}
\begin{barticle}[author]
\bauthor{\bsnm{Haslbeck},~\bfnm{Jonas~MB}\binits{J.~M.}} \AND
  \bauthor{\bsnm{Waldorp},~\bfnm{Lourens~J}\binits{L.~J.}}
(\byear{2018}).
\btitle{How well do network models predict observations? On the importance of
  predictability in network models}.
\bjournal{Behavior Research Methods}
\bvolume{50}
\bpages{853--861}.
\end{barticle}
\endbibitem

\bibitem[\protect\citeauthoryear{Holland and
  Rosenbaum}{1986}]{HollandRosenbaum1986}
\begin{barticle}[author]
\bauthor{\bsnm{Holland},~\bfnm{P.~W.}\binits{P.~W.}} \AND
  \bauthor{\bsnm{Rosenbaum},~\bfnm{P.~R.}\binits{P.~R.}}
(\byear{1986}).
\btitle{Conditional Association and Unidimensionality in Monotone Latent
  Variable Models}.
\bjournal{The Annals of Statistics}
\bvolume{14}
\bpages{1523--1543}.
\end{barticle}
\endbibitem

\bibitem[\protect\citeauthoryear{Hyv{\"a}rinen}{2006}]{Hyvarinen:2006}
\begin{barticle}[author]
\bauthor{\bsnm{Hyv{\"a}rinen},~\bfnm{Aapo}\binits{A.}}
(\byear{2006}).
\btitle{Consistency of pseudolikelihood estimation of fully visible Boltzmann
  machines}.
\bjournal{Neural Computation}
\bvolume{18}
\bpages{2283--2292}.
\end{barticle}
\endbibitem

\bibitem[\protect\citeauthoryear{Ip}{2001}]{Ip:2001}
\begin{barticle}[author]
\bauthor{\bsnm{Ip},~\bfnm{Edward~HS}\binits{E.~H.}}
(\byear{2001}).
\btitle{Visualizing multiple regression}.
\bjournal{Journal of Statistics Education}
\bvolume{9}.
\end{barticle}
\endbibitem

\bibitem[\protect\citeauthoryear{Junker and Ellis}{1997}]{Junker:1997}
\begin{barticle}[author]
\bauthor{\bsnm{Junker},~\bfnm{B.~W.}\binits{B.~W.}} \AND
  \bauthor{\bsnm{Ellis},~\bfnm{J.~L.}\binits{J.~L.}}
(\byear{1997}).
\btitle{A characterization of monotone unidimensional latent variable models}.
\bjournal{The Annals of Statistics}
\bvolume{25}
\bpages{1327-1343}.
\end{barticle}
\endbibitem

\bibitem[\protect\citeauthoryear{Kennedy}{2002}]{Kennedy:2002}
\begin{barticle}[author]
\bauthor{\bsnm{Kennedy},~\bfnm{Peter~E}\binits{P.~E.}}
(\byear{2002}).
\btitle{More on Venn diagrams for regression}.
\bjournal{Journal of Statistics Education}
\bvolume{10}.
\end{barticle}
\endbibitem

\bibitem[\protect\citeauthoryear{Lauritzen}{1996}]{Lauritzen96}
\begin{bbook}[author]
\bauthor{\bsnm{Lauritzen},~\bfnm{S.~L.}\binits{S.~L.}}
(\byear{1996}).
\btitle{Graphical Models}.
\bpublisher{Oxford University Press}, \baddress{Oxford}.
\end{bbook}
\endbibitem

\bibitem[\protect\citeauthoryear{Marsman et~al.}{2015}]{MarsmanEtAl2015SciRep}
\begin{barticle}[author]
\bauthor{\bsnm{Marsman},~\bfnm{M.}\binits{M.}},
  \bauthor{\bsnm{Maris},~\bfnm{{G. K. J. }}\binits{G.}},
  \bauthor{\bsnm{Bechger},~\bfnm{{T. M. }}\binits{T.}} \AND
  \bauthor{\bsnm{Glas},~\bfnm{{C. A. W. }}\binits{C.}}
(\byear{2015}).
\btitle{Bayesian Inference for Low-Rank {I}sing Networks}.
\bjournal{Scientific Reports}
\bvolume{5}.
\bdoi{10.1038/srep09050}
\end{barticle}
\endbibitem

\bibitem[\protect\citeauthoryear{Marsman et~al.}{2018}]{MarsmanEtAL_2018_MBR}
\begin{barticle}[author]
\bauthor{\bsnm{Marsman},~\bfnm{M.}\binits{M.}},
  \bauthor{\bsnm{Borsboom},~\bfnm{D.}\binits{D.}},
  \bauthor{\bsnm{Kruis},~\bfnm{J.}\binits{J.}},
  \bauthor{\bsnm{Epskamp},~\bfnm{S.}\binits{S.}}, \bauthor{\bsnm{{van
  Bork}},~\bfnm{R.}\binits{R.}}, \bauthor{\bsnm{Waldorp},~\bfnm{{L. J.
  }}\binits{L.}}, \bauthor{\bsnm{{van der Maas}},~\bfnm{{H. L. J.
  }}\binits{H.}} \AND \bauthor{\bsnm{Maris},~\bfnm{{G. K. J. }}\binits{G.}}
(\byear{2018}).
\btitle{An Introduction to Network Psychometrics: {R}elating {I}sing Network
  Models to Item Response Theory Models}.
\bjournal{Multivariate Behavioral Research}
\bvolume{53}
\bpages{15--35}.
\bdoi{10.1080/00273171.2017.1379379}
\end{barticle}
\endbibitem

\bibitem[\protect\citeauthoryear{Meinshausen and
  B\"{u}hlmann}{2006}]{Meinshausen:2006}
\begin{barticle}[author]
\bauthor{\bsnm{Meinshausen},~\bfnm{N.}\binits{N.}} \AND
  \bauthor{\bsnm{B\"{u}hlmann},~\bfnm{P.}\binits{P.}}
(\byear{2006}).
\btitle{High-dimensional graphs and variable selection with the Lasso}.
\bjournal{The Annals of Statistics}
\bvolume{34}
\bpages{1436-1462}.
\end{barticle}
\endbibitem

\bibitem[\protect\citeauthoryear{Nguyen}{2017}]{Nguyen:2017}
\begin{barticle}[author]
\bauthor{\bsnm{Nguyen},~\bfnm{H.~D.}\binits{H.~D.}}
(\byear{2017}).
\btitle{Near Universal Consistency of the Maximum Pseudolikelihood Estimator
  for Discrete Models}.
\bjournal{Annals of Statistics}
\bvolume{2}
\bpages{22-23}.
\end{barticle}
\endbibitem

\bibitem[\protect\citeauthoryear{Pearl}{2001}]{Pearl:2001}
\begin{barticle}[author]
\bauthor{\bsnm{Pearl},~\bfnm{J.}\binits{J.}}
(\byear{2001}).
\btitle{Causal Inference in the Health Sciences: A Conceptual Introduction}.
\bjournal{Health Services \& Outcomes Research Methodology}
\bvolume{2}
\bpages{189--220}.
\end{barticle}
\endbibitem

\bibitem[\protect\citeauthoryear{Schott}{1997}]{Schott97}
\begin{bbook}[author]
\bauthor{\bsnm{Schott},~\bfnm{J.~R.}\binits{J.~R.}}
(\byear{1997}).
\btitle{Matrix analysis for statistics}.
\bpublisher{John Wiley \& Sons}, \baddress{New York}.
\end{bbook}
\endbibitem

\bibitem[\protect\citeauthoryear{Tibshirani}{1996}]{Tibshirani:1996}
\begin{barticle}[author]
\bauthor{\bsnm{Tibshirani},~\bfnm{Robert}\binits{R.}}
(\byear{1996}).
\btitle{Regression shrinkage and selection via the lasso}.
\bjournal{Journal of the Royal Statistical Society. Series B (Methodological)}
\bpages{267--288}.
\end{barticle}
\endbibitem

\bibitem[\protect\citeauthoryear{Wainwright}{2009}]{Wainwright:2009}
\begin{barticle}[author]
\bauthor{\bsnm{Wainwright},~\bfnm{Martin~J}\binits{M.~J.}}
(\byear{2009}).
\btitle{Sharp thresholds for high-dimensional and noisy sparsity recovery
  using-constrained quadratic programming (Lasso)}.
\bjournal{Information Theory, IEEE Transactions on}
\bvolume{55}
\bpages{2183--2202}.
\end{barticle}
\endbibitem

\bibitem[\protect\citeauthoryear{Wainwright}{2019}]{Wainwright:2019}
\begin{bbook}[author]
\bauthor{\bsnm{Wainwright},~\bfnm{Martin~J}\binits{M.~J.}}
(\byear{2019}).
\btitle{High-dimensional statistics: A non-asymptotic viewpoint}
\bvolume{48}.
\bpublisher{Cambridge University Press}.
\end{bbook}
\endbibitem

\bibitem[\protect\citeauthoryear{Waldorp, Marsman and
  Maris}{2019}]{Waldorp:2019}
\begin{barticle}[author]
\bauthor{\bsnm{Waldorp},~\bfnm{Lourens}\binits{L.}},
  \bauthor{\bsnm{Marsman},~\bfnm{Maarten}\binits{M.}} \AND
  \bauthor{\bsnm{Maris},~\bfnm{Gunter}\binits{G.}}
(\byear{2019}).
\btitle{Logistic regression and Ising networks: prediction and estimation when
  violating lasso assumptions}.
\bjournal{Behaviormetrika}
\bvolume{46}
\bpages{49}.
\end{barticle}
\endbibitem

\end{thebibliography}
\end{document}